\documentclass[preprint2]{aastex}

\newcommand{\ms}{\mbox{m\,s$^{-1}$}}

\newcommand{\Msun}{\mbox{M$_{\odot}$}}

\newcommand{\Mjup}{\mbox{M$_{\rm Jup}$}}

\slugcomment{Submitted to ApJ}

\shorttitle{A Double Planetary System around HD 4732}
\shortauthors{Sato et al.}

\begin{document}


\title{A Double Planetary System around the Evolved Intermediate-Mass
Star HD 4732}


\author{Bun'ei Sato\altaffilmark{1},
Masashi Omiya\altaffilmark{1},
Robert A.~Wittenmyer\altaffilmark{2},
Hiroki Harakawa\altaffilmark{1},
Makiko Nagasawa\altaffilmark{1},
Hideyuki Izumiura\altaffilmark{3,4},
Eiji Kambe\altaffilmark{3},
Yoichi Takeda\altaffilmark{4,5},
Michitoshi Yoshida\altaffilmark{6},
Yoichi Itoh\altaffilmark{7},
Hiroyasu Ando\altaffilmark{5},
Eiichiro Kokubo\altaffilmark{4,5},
and
Shigeru Ida\altaffilmark{1}
}
\email{satobn@geo.titech.ac.jp}

\altaffiltext{1}{Department of Earth and Planetary Sciences, Tokyo Institute of
Technology, 2-12-1 Ookayama, Meguro-ku, Tokyo 152-8551, Japan}
\altaffiltext{2}{Department of Astrophysics, School of Physics, 
University of NSW, 2052, Australia}
\altaffiltext{3}{Okayama Astrophysical Observatory, National
  Astronomical Observatory of Japan, Kamogata,
  Okayama 719-0232, Japan}
\altaffiltext{4}{The Graduate University for Advanced Studies,
  Shonan Village, Hayama, Kanagawa 240-0193, Japan}
\altaffiltext{5}{National Astronomical Observatory of Japan, 2-21-1 Osawa,
   Mitaka, Tokyo 181-8588, Japan}
\altaffiltext{6}{Hiroshima Astrophysical Science Center, Hiroshima University,
  Higashi-Hiroshima, Hiroshima 739-8526, Japan}
\altaffiltext{7}{Nishi-Harima Astronomical Observatory, Center for Astronomy,
University of Hyogo, 407-2, Nishigaichi, Sayo, Hyogo
679-5313, Japan}




\begin{abstract}
We report the detection of a double planetary system orbiting around
the evolved intermediate-mass star HD 4732 from precise Doppler
measurements at Okayama Astrophysical Observatory (OAO) and
Anglo-Australian Observatory (AAO).
The star is a K0 subgiant with a mass of 1.7 $M_{\odot}$ and solar
metallicity. The planetary system is composed of two giant planets
with minimum mass of $m\sin i=2.4~M_{\rm J}$, orbital period of 360.2 d and
2732 d, and eccentricity of 0.13 and 0.23, respectively. Based on
dynamical stability analysis for the system, we set the upper limit
on the mass of the planets to be about 28 $M_{\rm J}$ ($i>5^{\circ}$)
in the case of coplanar prograde configuration.
\end{abstract}





\section{Introduction}\label{intro}

Precise radial-velocity measurements of stars have revealed
more than 500 extrasolar planets in the last 20 years.
Thanks to the increase in velocity precision and the duration
of the planet search programs, not only single planets but also
many multiple-planet systems have been discovered.
Around solar-type stars, including those systems with long-term
radial-velocity trends, $\sim$30--50 \% of stars with giant planets
are multiple-planet systems \citep{fischer:2001,wright:2007,wright:2009},
and systems comprising less massive planets such as super-Earths
and Neptune-mass planets are more abundant \citep{mayor:2011,
howard:2012, witt11a}.

The multiple-planet systems give us deep insights into the formation
and evolution of planetary systems \citep[e.g.][]{ford:2006}.
For example, evidence of planet-planet scattering events could
be preserved in the current orbital configurations of planets
\citep[e.g.][]{ford:2005}, and the prevalence of orbital
mean-motion resonances constitutes strong support
for differential convergent orbital migration having occurred
\citep[e.g.][]{lee:2002, kley:2004}.
Their statistical properties may also provide us hints about
the mechanisms of these processes; the rather uniform
semimajor axis distribution without the pileup of hot Jupiters
and the jump at 1 AU that are seen in single planets, and
systematically lower eccentricities compared to single
planets \citep{wright:2009}.

Studying the long-term orbital stability of multiple-planet systems
is also important. Since the best fit orbital parameters derived from
radial-velocity data do not necessarily guarantee that the resulting orbits
are stable over long periods of time, dynamical stability analysis
can check and further constrain the system parameters, such that the
solutions consistent with the data are stable
\citep[e.g.][]{lovis:2006,pepe:2007,niedzielski:2009a,
wright:2009,johnson:2011b, robertson:2012b, wittenmyer12b}.

Here we report the detection of a double giant-planet system
around the evolved intermediate-mass star HD 4732 (K0 IV,
$M=1.7~M_{\odot}$), one of the 300 targets of the
Okayama Planet Search Program \citep[e.g.][]{sato:2012}.
This discovery was facilitated by joint precise radial-velocity
observations made from the Anglo-Australian Observatory. 
Discoveries of planets and multiple-planet systems around such
intermediate-mass stars have been rapidly growing in number
during the past several years.
Although the number of known multiple-planet systems around
intermediate-mass stars is still small, it is interesting that
many of them appear to be in mean-motion resonances such as
24 Sex \citep[2:1;][]{johnson:2011b}, HD 102272
\citep[4:1;][]{niedzielski:2009a},
$\nu$ Oph \citep[6:1;][]{quirrenbach:2011, sato:2012},
and BD20+2457 \citep[3:2;][]{niedzielski:2009b}.
Although detailed dynamical studies are required to determine
whether these systems are truly in resonance \citep[e.g.][]
{robertson:2012a, wittenmyer12c}, the existence of
such systems suggests that the planetary orbital migration process
for intermediate-mass stars is similar to that for solar-type ones.

This paper is organized as follows. The stellar properties are
presented in section \ref{stpara} and the observations are described
in section \ref{obs}. Orbital solution are presented in section \ref{ana}
and results of line shape analysis are described in section \ref{line shape}.
Dynamical stability analysis is presented in section \ref{dynamics}
and section \ref{summary} is devoted to summary.

\section{Stellar Properties}\label{stpara}

HD 4732 (HR 228, HIP 3834) is listed in the Hipparcos catalog
\citep{esa:1997} as a K0 giant with the apparent $V$-band
magnitude $V=5.90$ and the parallax $\pi=17.70\pm0.99$ mas,
giving the distance of 56.5$\pm$3.2 pc and the absolute
visual magnitude $M_V=+2.04$ taking into account the correction of
interstellar extinction $A_V=0.10$ based on \citet{arenou:1992}'s
table \citep{takeda:2008}.

Atmospheric parameters (effective temperature $T_{\rm eff}$,
surface gravity $\log g$, micro-turbulent velocity $v_t$, and
Fe abundance [Fe/H]) of all the targets for Okayama Planet Search
Program were derived by \citet{takeda:2008} based on
the spectroscopic approach using the equivalent widths of
well-behaved Fe I and Fe II lines of iodine-free stellar spectra.
Details of the procedure and resultant parameters are presented
in \citet{takeda:2002} and \citet{takeda:2008}.
For HD 4732, they obtained $T_{\rm eff}=4959$ K, $\log g~(\rm{cgs}) =3.16$,
$v_t=1.12$ km s$^{-1}$, and [Fe/H]$=+$0.01 as well as projected
rotational velocity of $v\sin i_{\rm rot}=1.45$ km s$^{-1}$.

\begin{figure}
\plotone{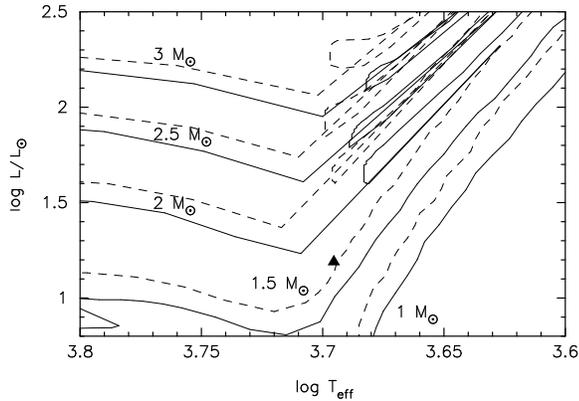}\epsscale{0.9}
\caption{HR diagram for HD 4732. Pairs of evolutionary tracks
from Lejeune and Schaerer (2001) for stars with $Z=0.02$
(solar metallicity; solid lines) and $Z=0.008$ (dashed lines)
of masses between 1 and 3 $M_{\odot}$ are also shown.}
\label{fig-HRD}
\end{figure}

With use of these atmospheric parameters together with $M_V$,
a bolometric correction based on the \citet{kurucz:1993}'s
theoretical calculation, and theoretical evolutionary tracks
of \citet{lejeune:2001}, \citet{takeda:2008} also determined
luminosity $L$ and mass $M$ for HD 4732 to be $L=15.5~L_{\odot}$
and $M=1.74~M_{\odot}$. The stellar radius $R$ was estimated to
be $R=5.4~R_{\odot}$ using the Stefan-Boltzmann relationship
and the measured $L$ and $T_{\rm eff}$. 
Although the star is listed in the Hipparcos catalog as a giant
star, the star is better regarded as a subgiant rather than giant
based on the surface gravity and its position on the HR diagram
(Fig \ref{fig-HRD}).
The stellar properties are summarized in Table \ref{tbl-stars}.

\begin{figure}
\plotone{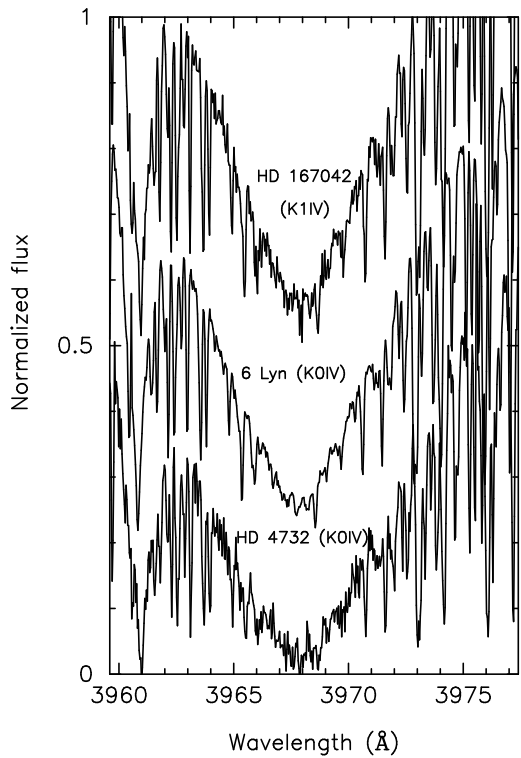}\epsscale{0.9}
\caption{Spectra in the region of Ca H lines. Stars with similar
spectral type to HD 4732 in our sample are also shown.
A vertical offset of about 0.3 is added to each spectrum.}
\label{fig-CaH}
\end{figure}

Hipparcos observations revealed photometric stability for the star
down to $\sigma_{\rm HIP}=0.006$ mag. Furthermore, the star shows no
significant emission in the core of Ca II HK lines as shown in
Figure \ref{fig-CaH}, which suggests that the star is
chromospherically inactive. For subgiants like HD 4732,
\cite{johnson:2010} reported a typical radial-velocity jitter,
which arises from a number of phenomena intrinsic to the star
such as granulation, oscillation, and activity,
of about 5 m s$^{-1}$. The value is comparable to that
of 6 m s$^{-1}$ estimated for HD 4732 in section \ref{ana}.

\section{Observation}\label{obs}

\subsection{OAO Observations}

Observations of HD 4732 at OAO were made with the 1.88 m telescope and
the HIgh Dispersion Echelle Spectrograph \citep[HIDES;][]{izumiura:1999}
from August 2004 to January 2012.
A slit width of the spectrograph was set to 200 $\mu$m ($0.76^{\prime\prime}$)
corresponding to a spectral resolution ($R=\lambda/\Delta\lambda$)
of 67000 by about 3.3 pixels sampling.
For precise radial-velocity measurements, we used an iodine absorption
cell \citep[I$_2$ cell;][]{kambe:2002}, which provides a fiducial
wavelength reference in a wavelength range of 5000--5800${\rm \AA}$.
We have obtained 48 data points of HD 4732 with signal-to-noise
ratio S/N$=$60--250 pix$^{-1}$ by an exposure time of 900--1800 sec
depending on the weather condition.
The reduction of echelle data (i.e. bias subtraction, flat-fielding,
scattered-light subtraction, and spectrum extraction) was performed
using the IRAF\footnote{IRAF is distributed by the National
Optical Astronomy Observatories, which is operated by the
Association of Universities for Research in Astronomy, Inc. under
cooperative agreement with the National Science Foundation,
USA.} software package in the standard way.

For radial-velocity analysis, we modeled I$_2$-superposed stellar spectra
(star+I$_2$) by the method detailed in \citet{sato:2002} and
\citet{sato:2012}, which is based on the method by
\citet{butler:1996} and \citet{val:95}.
In the method, we model a star+I$_2$ spectrum as a product of a
high resolution I$_2$ and a stellar template spectrum convolved
with a modeled point spread function (PSF) of the spectrograph.
We obtain the stellar spectrum by deconvolving a pure stellar
spectrum with the spectrograph PSF estimated from an I$_2$-superposed
B-star spectrum. We have achieved a long-term Doppler precision
of about 4--5 m s$^{-1}$ over a time span of 9 years. Measurement
error was derived from an ensemble of velocities from each
of $\sim$300 spectral chunks (each $\sim$3${\rm \AA}$ long)
in every exposure. We listed the derived radial velocities for OAO data
in Table~\ref{OAOvels} together with the estimated uncertainties.

\subsection{AAT Observations}

As HD\,4732 is near the southern limit for OAO ($\delta\sim-25^{\circ}$),
it is desirable to make 
observations from a more southerly site.  To improve the phase coverage 
for HD\,4732, in 2010 September we began observing it with the UCLES 
echelle spectrograph \citep{diego:90} at the 3.9-metre Anglo-Australian 
Telescope (AAT).  UCLES achieves a resolution of 45,000 with a 
1$^{\prime\prime}$ slit.  An iodine absorption cell provides wavelength 
calibration from 5000 to 6200\,\AA.  The spectrograph PSF
and wavelength calibration are derived from the iodine 
absorption lines embedded on every pixel of the spectrum by the cell 
\citep{val:95,butler:1996}. The result is a precision Doppler velocity 
estimate for each epoch, along with an internal uncertainty estimate, 
which includes the effects of photon-counting uncertainties, residual 
errors in the spectrograph PSF model, and variation in the underlying 
spectrum between the iodine-free template, and epoch spectra observed 
through the iodine cell.  The photon-weighted mid-time of each exposure 
is determined by an exposure meter.  This technique has been 
successfully used at the AAT by the Anglo-Australian Planet Search (e.g. 
Tinney et al. 2001, O'Toole et al. 2009, Jones et al. 2010, Tinney et 
al. 2011a) and the Pan-Pacific Planet Search \citep{wittenmyer:2011}.  All 
velocities are measured relative to the zero-point defined by the 
template observation.  AAT/UCLES precision velocities are obtained using 
the \textit{Austral} code \citep{endl00}.

We have obtained 19 AAT observations of HD\,4732, and an iodine-free 
template observation was obtained on 2010 Oct 25.  Exposure times ranged 
from 400 to 1200 s, with a resulting S/N of $\sim$100-200 per pixel each 
epoch.  The AAT data span a total of 636 days, and have a mean internal 
velocity uncertainty of 5.0\,\ms.  The data are given in 
Table~\ref{AATvels}.

\section{Orbit Fitting and Planetary Parameters}\label{ana}

For candidate multiple-planet systems, and for planet candidates with 
orbital periods near one year, it is critical to obtain the most 
complete possible phase coverage, and it is ideal to independently 
confirm the signal(s) from independent observatories.  A recent example 
is HD\,38283b, a 0.34\Mjup\ planet with a period $P=363.2\pm$1.6 days 
\citep{tinney11b}.  That planet required 12 years of observations to 
confirm, and a robust detection was further aided by the fact that 
HD\,38283 is circumpolar and so could be observed year-round (although 
at high airmasses below the pole).  Another example is HD\,159868c, 
which has a period $P=352.3\pm$1.3 days \citep{wittenmyer12b}, also 
worryingly close to one year.  Again, more than 9 years of AAT data were 
needed for a secure detection, and the 352-day period was confirmed 
independently with Keck data over a 4-year span \citep{wittenmyer12b}.

Five years of radial-velocity observations of HD\,4732 at OAO
revealed evidence of a signal with a period near one year
($\sim$338 days).  Since this 
star is near the southern limit for OAO, it is only observable for 5 
months of the year, resulting in persistent phase gaps which make the 
confirmation of such a period extremely difficult.  AAT observations in 
2010-2011 filled in the critical phase gaps and confirmed the orbital 
period suggested by the OAO data.  Continued observations from both 
telescopes in 2011-2012 have also revealed a second velocity signal, 
first manifesting as a residual trend, and subsequently as a second, 
long periodicity.

To fit the two Keplerian signals evident in the HD\,4732 data, we first 
employed a genetic algorithm.  This approach has proven useful in 
previous work where it was necessary to fit highly uncertain orbits with 
long periods near the total duration of observations \citep{HUAqr2, 
wittenmyer12b, NNSer}.  We ran the genetic algorithm for 10,000 
iterations, each consisting of ~1000 individual trial fits.  The 
best-fit set of parameters is thus the result of $\sim\,10^7$ trial 
fits.  The parameters of the best 2-planet solution obtained by the 
genetic algorithm were then used as initial inputs for the GaussFit code 
\citep{jefferys87}, a nonlinear least-squares fitting routine.  The 
GaussFit model has the ability to allow the offsets between data sets to 
be a free parameter. This is important because the radial velocities 
from OAO and AAT are not absolute radial velocities, but rather are 
measured relative to the iodine-free stellar template.  Each data set 
thus has an arbitrary zero-point offset which must be accounted for in 
the orbit-fitting procedure \citep{witt09}.  These offsets are 
3.8$\pm$3.2 \ms\ (OAO) and $-$0.1$\pm$3.7 \ms\ (AAT).

For the final orbit fitting, we added radial-velocity jitter in 
quadrature to the velocity uncertainty of each observation.  This jitter 
arises from a number of phenomena intrinsic to the star, such as 
granulation, oscillations, and activity \citep{saar98,wright05}.  As 
HD\,4732 is an evolved star, scaling relations \citep{kb95, kb11} are 
not applicable.  We thus determine the appropriate level of jitter 
empirically, by performing two-Keplerian fits with varying levels of 
jitter.  On some dates, the AAT recorded multiple consecutive exposures 
of HD\,4732 (Table~\ref{AATvels}).  This provides one way to estimate 
the jitter: by noting the velocity spread of the multiple exposures.  On 
the four dates which had multiple exposures, the mean velocity spread is 
2.9\ms.  This is, however, a lower limit as the timescales of many 
jitter sources are on the order of hours (granulation) or months 
(rotation).  We also try a range of jitters from 4--10 \ms, and note the 
results of the fits in Table~\ref{jitfits}.  The estimation of 
radial-velocity jitter is quite uncertain, and this is reflected in 
Table~\ref{jitfits}, where we see that the amount of added jitter has 
little effect on the quality of the resulting fit.  We adopt a jitter 
estimate of 6\ms, as this value brings the reduced $\chi^2$ of the fit 
close to unity.  The resulting 2-Keplerian planetary parameters are 
given in Table~\ref{planetparams}.  The uncertainties on the fitted 
parameters are derived in the usual way, from the covariance matrix of 
the least-squares fit.  The phase gap for planet b has the greatest 
impact on the uncertainty in $T_0$ for that planet.  As a result, the 
uncertainty $\sigma_T$=18.4 days remains large even though many orbital 
cycles of planet b have been observed.  The two-planet fit has a reduced 
$\chi^2$ of 1.05, and the RMS scatter about the fit is 7.09\ms.  Using a 
stellar mass of $1.74^{+0.14}_{-0.20}$ \Msun \citep{takeda:2008}, the 
planets have minimum masses m~sin~$i$ of 2.4$\pm$0.3 (inner planet) and 
2.4$\pm$0.4 \Mjup (outer planet).

\section{Line Shape Analysis}\label{line shape}

In order to investigate other possible causes of observed radial-velocity
variations such as pulsation and rotational modulation rather than orbital motion,
spectral line shape analysis was performed with use of high resolution
stellar template spectra.  Details of the analysis method are described
in \citet{sato:2007} and \citet{sato:2002}.

\begin{figure}\epsscale{0.9}
\plotone{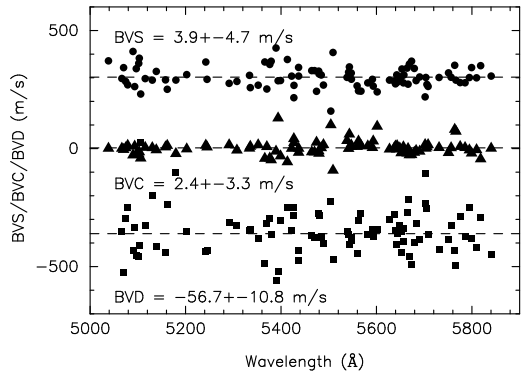}
\caption{Bisector quantities of cross-correlation profiles between
the templates of HD 4732 at phases with $\sim$ 40 m s$^{-1}$ and
with $\sim-$ 20 m s$^{-1}$ of the observed radial velocities:
bisector velocity span (BVS, circles),
bisector velocity curvature (BVC, triangles), and bisector velocity
displacement (BVD, squares). Offsets of 300 m s$^{-1}$ and
$-$300 m s$^{-1}$ are added to BVS and BVD, respectively.}
\label{fig:bvs}
\end{figure}

At first, two stellar templates were extracted from five star+I$_2$ spectra
at phases with different radial-velocity level, $\sim$ 40 m s$^{-1}$ and
$\sim-$20 m s$^{-1}$, of the observed radial velocities based on the method
by \citet{sato:2002}.
Cross correlation profiles of the two templates were calculated
for about 90 spectral chunks (4--5${\rm \AA}$ width each) in
which severely blended lines or broad lines were not included,
and then three bisector quantities were derived for the
cross correlation profile for each chunk:
the velocity span (BVS), which is the velocity difference
between two flux levels of the bisector;
the velocity curvature (BVC), which is the difference of the
velocity span of the upper half and lower half of the bisector;
and the velocity displacement (BVD), which is the average of
the bisector at three different flux levels.
Here we used flux levels of 25\%, 50\%, and 75\% of the cross
correlation profile to calculate the above three
bisector quantities, and obtained BVS$=$3.9$\pm$4.7 m s$^{-1}$,
BVC$=$2.4$\pm$3.3 m s$^{-1}$, and BVD$=-$57$\pm$11 m s$^{-1}$ for HD 4732
(Figure \ref{fig:bvs}).
Both of the BVS and the BVC are identical to zero and the
average BVD agrees with the velocity difference between the two
templates. Then the cross correlation profiles can be considered to be
symmetric and thus the observed radial-velocity variations are best
explained by parallel shifts of the spectral lines rather than
distortion of them, which is consistent with the orbital-motion hypothesis.

\section{Dynamical Stability Analysis}\label{dynamics}

\begin{figure}\epsscale{0.9}
\plotone{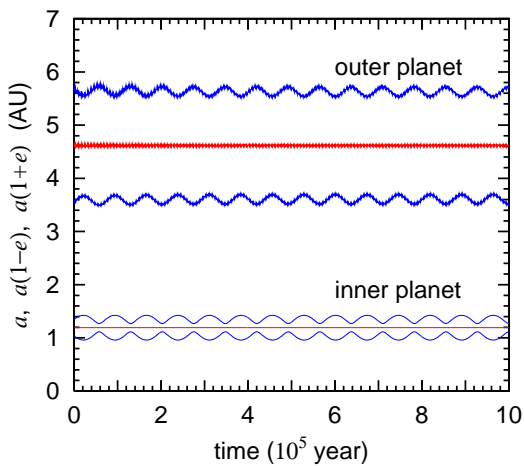}
\caption{Evolution of semimajor axis $a$, pericenter $a(1-e)$,
and apocenter $a(1+e)$ distances for the two planets of HD 4732 system for
the best-fit two-Keplerian model to the radial-velocity data. Red lines
are for semimajor axis and blue ones are for pericenter (lower
line for each planet) and apocenter (upper line for each planet) distance.
We here assume $i=90^{\circ}$ and prograde coplanar orbits.  
\label{fig:orbitcalc}}
\end{figure}

We here present dynamical studies of the HD 4732 system.
Since the eccentricities of the planets are relatively small and
it is thus considered that they probably have not experienced
close encounters, we here assume that the both planets share
the same orbital plane and are prograde.
For the numerical integrations, we use a 4-th order Hermite scheme. 
Figure~\ref{fig:orbitcalc} shows the one million year evolution of
the best-fit two-Keplerian model derived in section \ref{ana},
for $i=90^{\circ}$ orbits (i.e.~planet masses at minimum). 
The system has two secular eigenfrequencies related with eccentricity,
11$^{\prime\prime}$/yr and 28$^{\prime\prime}$/yr. Accordingly, the
planetary pericenter, $a(1-e)$, and apocenter, $a(1+e)$, distances
oscillate with the secular timescale of 76000 yr.

\begin{figure}\epsscale{0.9}
\plotone{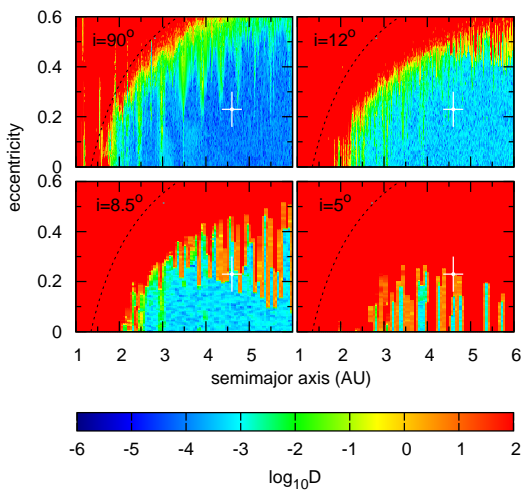}
\caption{Stability map of the outer planet (HD 4732 c) for a coplanar
prograde configuration with different values of inclination $i$.
Orbital elements other than semimajor axis and
eccentricity of the outer planet are taken from the best fit values. 
The color scale shows the level of mean motion diffusion, $\log_{10} D$. 
Blue to navy orbits are stable, while yellow to red orbits are chaotic.
The step size of eccentricity is $\Delta e$=0.005. The step sizes of
semimajor axis are $\Delta a$=0.005 AU (top-left panel),  $\Delta a$=0.01 AU
(top-right panel), and $\Delta a$=0.05 AU (bottom panels).
White crosses represent the best fitted $a$ and $e$ with
their 1 $\sigma$ errors. Black dotted line is the orbit crossing line. 
\label{fig:stablmap}}
\end{figure}

For lower inclinations, since the mass increases
proportional to $1/\sin i$, mutual perturbations become stronger. 
The stability map in the ($a, e$) plane of the outer planet for different
values of $i$ is shown in Fig.~\ref{fig:stablmap}.
Here we plot the stability index
$D=|\langle n_2 \rangle -\langle n_1 \rangle |$ (in $^{\circ}$/yr)
following \cite{couetdic:2010}, which studied stability of HD202206
system using Laskar's frequency map analysis \citep{laskar:1990, laskar:2001}. 
Following the analysis, we calculate an average of mean motion of
the outer planet in 1000 Kepler periods ($\langle n_1 \rangle$) and subtract it
from an average of mean motion of the same planet obtained
in the next consecutive 1000 Kepler periods ($\langle n_2 \rangle$). 
Through comparison with the five times longer simulations, we find that
the system is regular when $\log_{10} D \le -3$ (blue to navy), while
the system is chaotic when $\log_{10} D>-1$ (yellow to red). 
As inclination decreases, the system loses stability. There is no
stable area around 1$\sigma$ of the best fitted coplanar orbits
for $i \le 5^{\circ}$. Thus, in the coplanar prograde configuration,
we can roughly set the upper limit on the mass of both of the two
planets to be about $28 M_{\rm Jup}$ $(i > 5^{\circ})$, which falls in the
substellar mass regime.

In top left panel of Fig.~\ref{fig:stablmap}, one can find effects of
mean-motion resonances as green vertical incisions. In the case
of this system, based on the best-fit values of argument of pericenter,
orbits closer to the mean-motion resonances are more unstable.
The best-fit solution derived by the radial-velocity data (white cross) sits
between the 7:1 and 8:1 mean-motion resonances.
From the orbital integration shown in Fig.~\ref{fig:orbitcalc},
we have also confirmed that their resonant angles are not librating. 
We thus conclude that the outer planet is not in the mean-motion
resonances of the inner planet.

\section{Summary}\label{summary}

We report the detection of a double planet system around
the evolved intermediate-mass star HD 4732 (K0 IV, $M=1.7~M_{\odot}$)
by precise radial-velocity measurements at OAO and AAO. The system is composed
of two giant planets with minimum masses of $m_2\sin i=2.4~M_{\rm J}$,
orbital period of 360.2 d and 2732 d, and eccentricity of 0.13 and 0.23,
respectively. The joint observations of OAO and AAO allowed us
to increase the phase coverage of the nearly 1-yr periodicity of the
inner planet and detect long-period signal of the second outer planet.

The configuration of the system is similar to those of other multiple-planet
systems currently known: two giant planets in a combination
of intermediate- and long-period orbits with relatively low eccentricities.
HD 4732 c is the outer most planetary candidate ($m_p\sin i<13~M_{\rm J}$)
ever discovered around
intermediate-mass stars except for those discovered by direct imaging
\footnote{$\nu$ Oph c \citep[$m\sin i=27~M_{\rm J}$, a=6.1 AU;][]{sato:2012}
is the outer most brown-dwarf companion ever discovered around
intermediate-mass stars by radial velocity measurements.}.
The period ratio of the two planets is close to 7:1 or 8:1, but dynamical
analysis showed that the system is not in the mean-motion resonance
and orbits close to mean-motion resonances are unstable in this case.

One million year orbital evolution of the best-fit two-Keplerian
model in a coplanar prograde configuration showed no significant
changes in eccentricities, but showed their slight oscillations
with secular timescale of 76000 yr. Based on the dynamical stability
analysis, we found that the system is dynamically unstable in the
case of $i\le5^{\circ}$ for a coplanar prograde configuration.
Thus, we can set the upper limit on the mass of both of the
two planets to be about 28 $M_{\rm J}$ ($i>5^{\circ}$) in this case,
which falls well into the substellar mass regime.


\acknowledgments

This research is based on data collected at Okayama Astrophysical 
Observatory (OAO), which is operated by National Astronomical 
Observatory of Japan, and at the Anglo-Australian Observatory. We are 
grateful to all the staff members of OAO for their support during the 
observations. We thank students of Tokyo Institute of Technology and 
Kobe University for their kind help for the observations. BS was partly 
supported by MEXT's program "Promotion of Environmental Improvement for 
Independence of Young Researchers" under the Special Coordination Funds 
for Promoting Science and Technology, and by Grant-in-Aid for Young 
Scientists (B) 20740101 from the Japan Society for the Promotion of 
Science (JSPS). RW is supported by a UNSW Vice-Chancellor's Fellowship. 
MN is supported by Grant-in-Aid for Young Scientists (B) 21740324 and HI 
is supported by Grant-In-Aid for Scientific Research (A) 23244038 from 
JSPS.

This research has made use of the SIMBAD database, operated at
CDS, Strasbourg, France.


\clearpage

\begin{deluxetable}{lr}
\tabletypesize{\small}
\tablecaption{Stellar Parameters for HD 4732\label{tbl-stars}}
\tablewidth{0pt}
\tablehead{
\colhead{Parameter} & \colhead{Value}}
\startdata
Spectral Type & K0 IV$^a$\\
$\pi$ (mas)     & 17.70$\pm$0.99\\
$V$             & 5.90\\
$B-V$           & 0.944\\
$A_V$            & 0.10\\
$M_V$            & 2.04\\
$B.C.$          & $-$0.27\\
$T_{\rm eff}$ (K) & 4959$\pm$25$^b$\\
$\log g$ (cgs) & 3.16$\pm$0.08$^b$\\
$v_{\rm t}$ (km s$^{-1}$) & 1.12$\pm$0.07$^b$\\
$[$Fe/H$]$ (dex)       & 0.01$\pm$0.04$^b$\\
$L$ ($L_{\odot}$) & 15.5\\
$R$ ($R_{\odot}$) & 5.4 (5.0--5.8)$^c$\\
$M$ ($M_{\odot}$) & 1.74 (1.60--1.94)$^c$\\
$v\sin i_{\rm rot}$ (km s$^{-1}$) & 1.45\\
$\sigma_{\rm HIP}$ (mag) & 0.006\\
\enddata
\tablenotetext{a}{The star is listed in the Hipparcos catalogue as a
K0 giant. But judged from the position of the star on the
HR diagram (Figure \ref{fig-HRD}), the star should be better
classified as a less evolved subgiant.}
\tablenotetext{b}{
The uncertainties of these values are internal statistical errors
(for a given data set of Fe~{\sc i} and Fe~{\sc ii} line equivalent
widths) evaluated by the procedure described in subsection 5.2 of
\cite{takeda:2002}.}
\tablenotetext{c}{
The values in the parenthesis correspond to
the range of the values assuming the realistic uncertainties in
$\Delta\log L$ corresponding to parallax errors in the Hipparcos
catalog, $\Delta\log T_{\rm eff}$ of $\pm0.01$ dex ($\sim\pm100$~K),
and $\Delta{\rm [Fe/H]}$ of $\pm0.1$ dex.}
\end{deluxetable}

\begin{deluxetable}{lrr}
\tabletypesize{\scriptsize}
\tablecolumns{3}
\tablewidth{0pt}
\tablecaption{OAO Radial Velocities for HD 4732}
\tablehead{
\colhead{JD-2400000} & \colhead{Velocity (\ms)} & \colhead{Uncertainty
(\ms)}}
\startdata
\label{OAOvels}
53249.25552  &    -55.9  &    4.7  \\
53333.01997  &    -11.6  &    3.7  \\
53579.26965  &    -57.7  &    4.3  \\
53636.21690  &    -69.4  &    6.4  \\
53660.13539  &    -62.8  &    4.1  \\
53693.98447  &    -30.2  &    3.8  \\
53728.95697  &     -4.8  &    8.1  \\
53730.93188  &    -14.4  &    4.3  \\
53744.94607  &     -3.3  &    4.1  \\
53963.26781  &    -64.7  &    3.8  \\
54018.11973  &    -37.6  &    3.8  \\
54049.02581  &    -17.8  &    4.4  \\
54088.93606  &     -8.3  &    3.3  \\
54114.88858  &      3.0  &    6.2  \\
54305.29326  &    -47.9  &    3.8  \\
54338.23288  &    -38.5  &    4.4  \\
54380.15468  &    -22.6  &    4.1  \\
54416.02081  &    -18.8  &    4.8  \\
54460.93793  &     24.1  &    3.0  \\
54495.89851  &     33.7  &    9.5  \\
54496.90370  &     46.0  &    4.9  \\
54703.25119  &    -38.0  &    3.6  \\
54756.10553  &     -7.5  &    3.5  \\
54796.00048  &     23.5  &    3.8  \\
54817.94637  &     31.3  &    3.5  \\
55101.16674  &     -5.9  &    4.0  \\
55108.20638  &      2.7  &    4.6  \\
55133.10162  &     11.9  &    3.8  \\
55137.09148  &      2.9  &    4.1  \\
55161.03266  &     32.6  &    3.7  \\
55162.00982  &     34.0  &    2.8  \\
55183.92889  &     42.0  &    3.6  \\
55203.89392  &     46.3  &    3.5  \\
55233.90048  &     68.5  &    5.8  \\
55399.28930  &    -13.2  &    3.7  \\
55438.17586  &    -22.0  &    4.4  \\
55444.21320  &    -29.9  &    3.5  \\
55468.16146  &      0.7  &    5.0  \\
55480.13068  &     13.1  &    4.3  \\
55503.02690  &     28.3  &    3.9  \\
55510.06421  &     22.8  &    3.7  \\
55524.97523  &     39.5  &    3.4  \\
55546.92022  &     41.9  &    3.6  \\
55579.91720  &     53.6  &    6.0  \\
55787.29129  &    -14.6  &    4.8  \\
55877.01304  &      6.5  &    3.5  \\
55919.93294  &     29.2  &    4.0  \\
55937.88506  &     41.8  &    3.4  \\
\enddata
\end{deluxetable}

\begin{deluxetable}{lrr}
\tabletypesize{\scriptsize}
\tablecolumns{3}
\tablewidth{0pt}
\tablecaption{AAT Radial Velocities for HD 4732}
\tablehead{
\colhead{JD-2400000} & \colhead{Velocity (\ms)} & \colhead{Uncertainty
(\ms)}}
\startdata
\label{AATvels}
55455.16094  &    -19.7  &    5.6  \\
55455.20587  &    -20.5  &    5.9  \\
55456.06022  &    -24.3  &    5.5  \\
55494.91653  &     21.9  &    4.0  \\
55579.92330  &     63.1  &    4.8  \\
55600.90903  &     67.7  &    5.9  \\
55600.91897  &     75.7  &    5.1  \\
55600.93330  &     71.8  &    5.3  \\
55707.30755  &     -3.1  &    4.3  \\
55707.31526  &     -4.4  &    4.4  \\
55760.15572  &    -36.3  &    4.0  \\
55783.13308  &    -37.3  &    6.9  \\
55783.14079  &    -35.9  &    5.8  \\
55841.99044  &    -12.3  &    4.3  \\
55878.95827  &      1.0  &    4.4  \\
55880.01556  &     10.3  &    4.5  \\
56060.29762  &    -31.2  &    5.6  \\
56089.30186  &    -39.6  &    4.2  \\
56091.27740  &    -46.8  &    4.2  \\
\enddata
\end{deluxetable}

\begin{deluxetable}{lrr}
\tabletypesize{\scriptsize}
\tablecolumns{3}
\tablewidth{0pt}
\tablecaption{Jitter Estimates}
\tablehead{
\colhead{Added jitter (\ms)} & \colhead{$\chi^2_{\nu}$} & \colhead{RMS (\ms)}}
\startdata
\label{jitfits}
2.9 & 2.12 & 7.10 \\
4.0 & 1.66 & 7.09 \\
5.0 & 1.32 & 7.09 \\
6.0 & 1.05 & 7.09 \\
7.0 & 0.85 & 7.08 \\
8.0 & 0.70 & 7.08 \\
9.0 & 0.59 & 7.08 \\
10.0 & 0.49 & 7.08 \\
\enddata
\end{deluxetable}

\begin{deluxetable}{lr@{$\pm$}lr@{$\pm$}lr@{$\pm$}lr@{$\pm$}lr@{$\pm$}lr@{$\pm$}
lr@{$\pm$}l}
\tabletypesize{\scriptsize}
\tablecolumns{10}
\tablewidth{0pt}
\tablecaption{Keplerian Orbital Solutions }
\tablehead{
\colhead{Planet} & \multicolumn{2}{c}{Period} & \multicolumn{2}{c}{$T_0$}
&
\multicolumn{2}{c}{$e$} & \multicolumn{2}{c}{$\omega$} &
\multicolumn{2}{c}{K } & \multicolumn{2}{c}{m sin $i$ } &
\multicolumn{2}{c}{$a$ } \\
\colhead{} & \multicolumn{2}{c}{(days)} & \multicolumn{2}{c}{(JD-2400000)}
&
\multicolumn{2}{c}{} &
\multicolumn{2}{c}{(degrees)} & \multicolumn{2}{c}{(\ms)} &
\multicolumn{2}{c}{(\Mjup)} & \multicolumn{2}{c}{(AU)}
 }
\startdata
\label{planetparams}   
HD 4732 b & 360.2 & 1.4 & 54967 & 18 & 0.13 & 0.06 & 85 & 16 &
47.3 & 3.5 & 2.37 & 0.34 & 1.19 & 0.05 \\
HD 4732 c & 2732 & 81 & 56093 & 103 & 0.23 & 0.07 & 118 & 15 &
24.4 & 2.2 & 2.37 & 0.38 & 4.60 & 0.23 \\
\enddata
\end{deluxetable}

\begin{figure}
\plotone{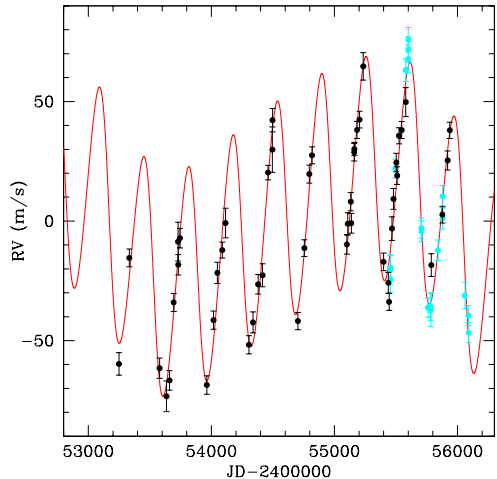}
\caption{Two-planet fit for HD\,4732 radial velocities. The planets have 
periods of 360 and 2732 days, and the RMS about this fit is 7.09\ms. OAO 
data are shown as filled black circles, and AAT data are filled cyan 
circles. }
\label{bothplot}
\end{figure}

\begin{figure}
\plottwo{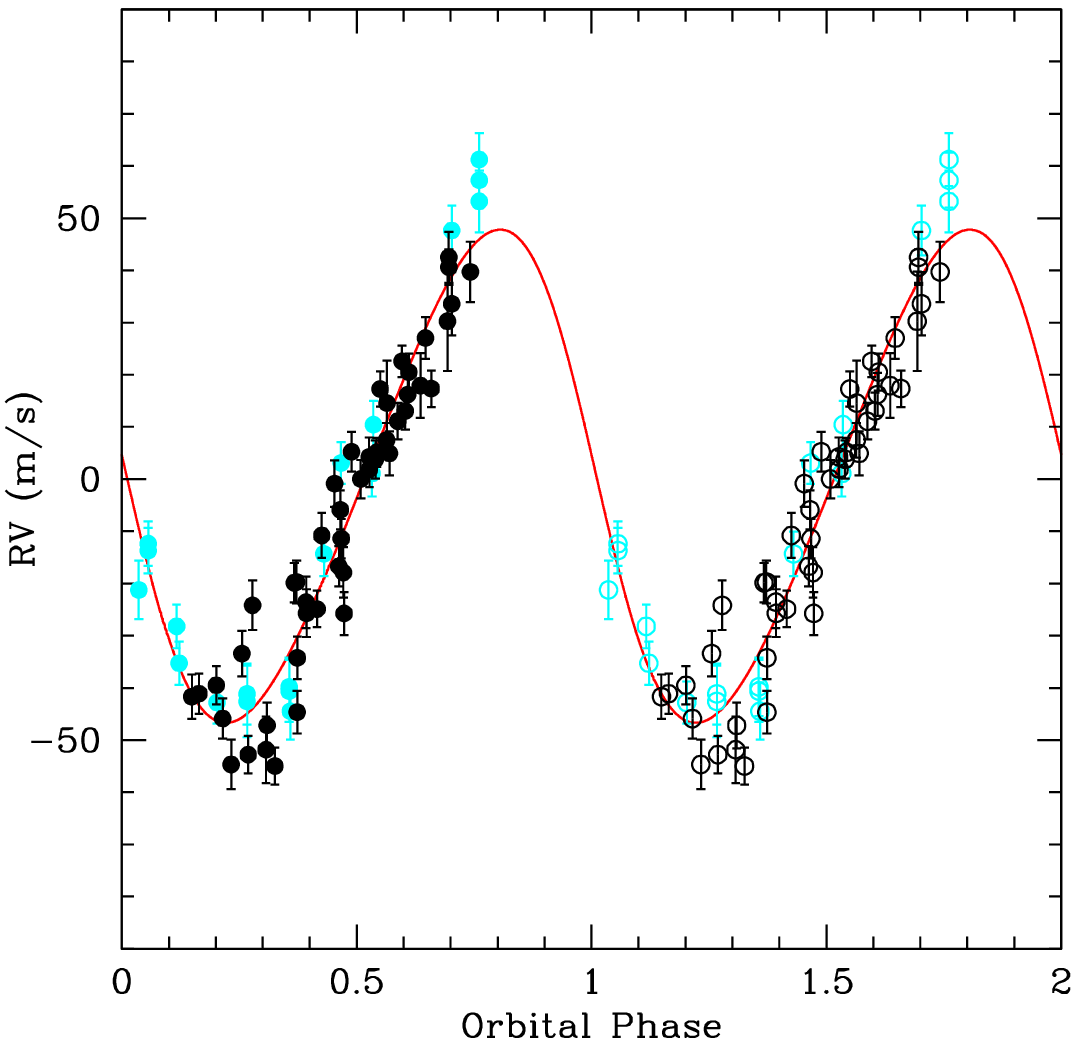}{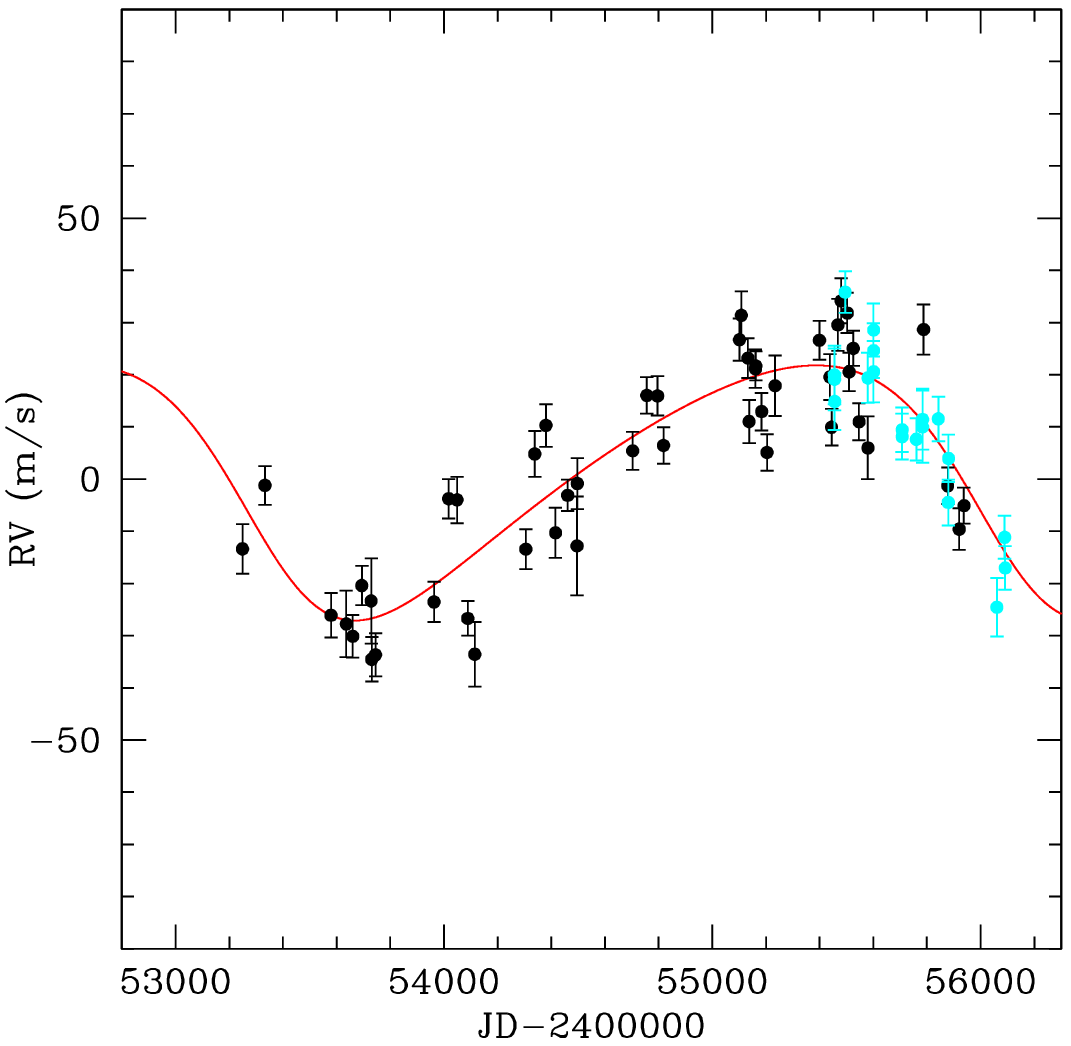}
\caption{Left: Phase plot for HD\,4732b ($P=360$ days), after removing 
the signal of the outer planet.  Two cycles are shown for clarity. 
Right: Radial-velocity observations for HD\,4732c ($P=2732$ days), after 
removing the signal of the inner planet.  The symbols have the same 
meaning as in Figure~\ref{bothplot}.}
\label{eachplanet}
\end{figure}

\end{document}